\documentclass[twocolumn,showpacs,pre,amsmath,amssymb,superscriptaddress,nofootinbib]{revtex4}
\usepackage{amsmath,amssymb,bm,graphicx,epsfig,ulem,color,natbib}
\newcommand{\bx}{\mathbf {x}}

\newcommand{\bv}{\mathbf {v}}
\newcommand{\bvf}{\mathbf{v}_{f}}
\newcommand{\etal}{{\it et al.~}}
\begin{document}

\title{\bf Fighting the flow: the stability of model flocks in a vortical flow}

\author{A.~W.~Baggaley}
\affiliation{School of Mathematics and Statistics, Newcastle
University, Newcastle upon Tyne, NE1 7RU, UK}\affiliation{Joint Quantum Centre Durham-Newcastle}
\begin{abstract}
We investigate the stability of self-propelled particle flocks in the Taylor-Green vortex, a steady vortical flow. We consider a model where particles align themselves to a combination of the orientation and the acceleration of particles within  a critical radius. We identify two distinct regimes, if alignment with orientation is dominant the particles tend to be expelled from regions of high vorticity. In contrast if anticipation is dominant the particles accumulate in areas of large vorticity. In both regimes the relative order of the flock is reduced. However we show that there can be a critical balance of the two effects which stabilises the flock in the presence of external fluid forcing. This strategy could provide a mechanism for animal flocks to remain globally ordered in the presence of fluid forcing, and may also have applications in the design of flocking autonomous drones and artificial microswimmers.
\end{abstract}
\maketitle

\section*{Introduction}\label{sec:intro}
In a vast range of biological systems, from bird flocks to fish schools to insect swarms, collective behaviour is observed. Studying why and how such collective behaviour arises can be important to first understand and then address a number of ecological issues, mainly due to human impact on the environment. In addition there are also important technological applications, collective robot motion for example \cite{Floreano:2015aa}. 

In this paper we investigate one of the most important and interesting examples of collective behaviour, collective motion. Whilst various modelling approaches have been suggested in the literature, one of the most popular is based on  self-propelled particles (SPPs), building on the seminal Vicsek model \cite{Vicsek_1995}. In this numerical approach $N$ particles move in a two dimensional domain (extension to higher dimensions is straighforward) with a constant velocity $V$. A particles direction of motion is instantaneously updated at every numerical time-step to align with neighbouring particles within some fixed critical radius, $R$. Noise is introduced in the system by applying a random rotation of a given size to each particle after the alignment step. This is to model intrinsic noise, due to the fact that animals will never perfectly align, and extrinsic noise, i.e. forcing from the external environment.

The number of subsequent variants of the Vicsek model is far too great to list here and we recommend the interested reader consult \cite{Vicsek201271} and references therein.
Whilst it has been shown that the behaviour of marching locusts could be modelled using an SPP approach \cite{Buhl2006}, Khurana \& Ouellette \cite{Khurana2013} showed that Vicsek flocks were particularly sensitive to spatio-temporally correlated noise. In particular flocks were more easily destabilised when the extrinsic noise consisted of a model of a turbulent flow, in contrast to the case where a random (delta-correlated in space and time) field forced the system.
Furthermore we recently showed \cite{Baggaley2015} that Vicsek flocks in a steady vortical flow are concentrated into areas of high vorticity. This has a profound effect on the morphology of the flock, with a dramatic increase in the filamentarity, i.e. the perimeter of the flock is increased for a given area. One reasons animals exhibit collective motion is it gives them a better chance of avoiding predation \cite{krause2002living}. If one assumes a predator generally will attack the closest individual, an animal can reduce the area (volume) of the region in which it is the closest prey to a predator by joining a `flock' \cite{Hamilton1971295}. Of course the size of this `domain of danger' is also dependent on the shape of the flock, with safety reducing if the perimeter (surface area) of a flock increases for a given area (volume). Hence our earlier findings \cite{Baggaley2015}  could have profound implications for animals flocking in a turbulent environment, or more likely animals have developed strategies to counteract this effect. Finding such a strategy is the goal of this paper, in particular (motivated by the recent study of Morin \etal \cite{Morin2015}), we wish to understand if both alignment and anticipation can stabilise model flocks in the presence of spatially correlated extrinsic noise.
%----------------------------------------------------------------------------------------------------------------------------------------
%\begin{overpic}[width=0.45\textwidth]{fig1a}
% \put (0,100) {\Large (a)}
%\end{overpic}

\begin{figure*}[ht]
\begin{center}
\includegraphics[width = 0.4\textwidth]{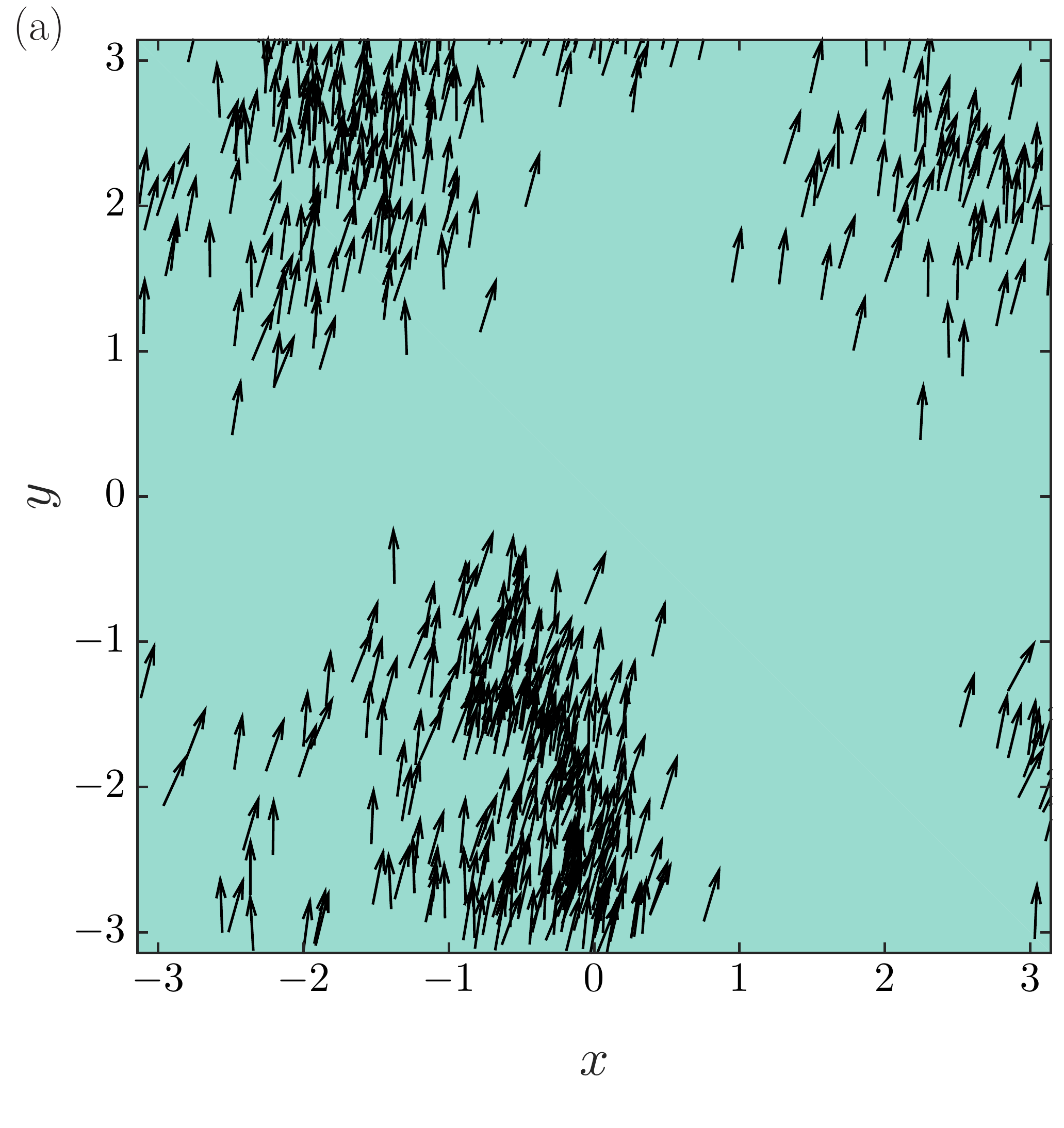}\hspace{5mm}
\includegraphics[width = 0.4\textwidth]{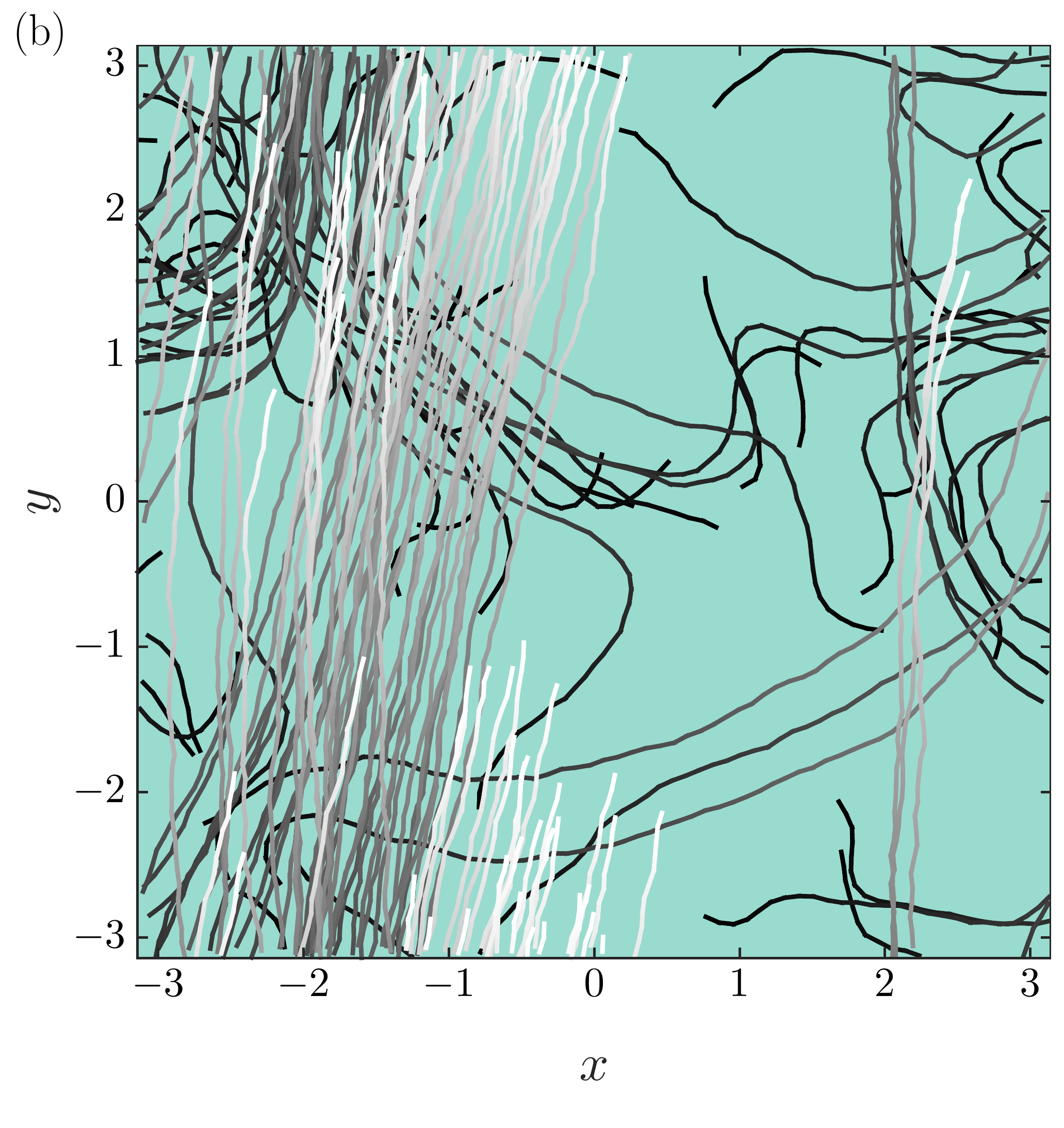}\\
\includegraphics[width = 0.4\textwidth]{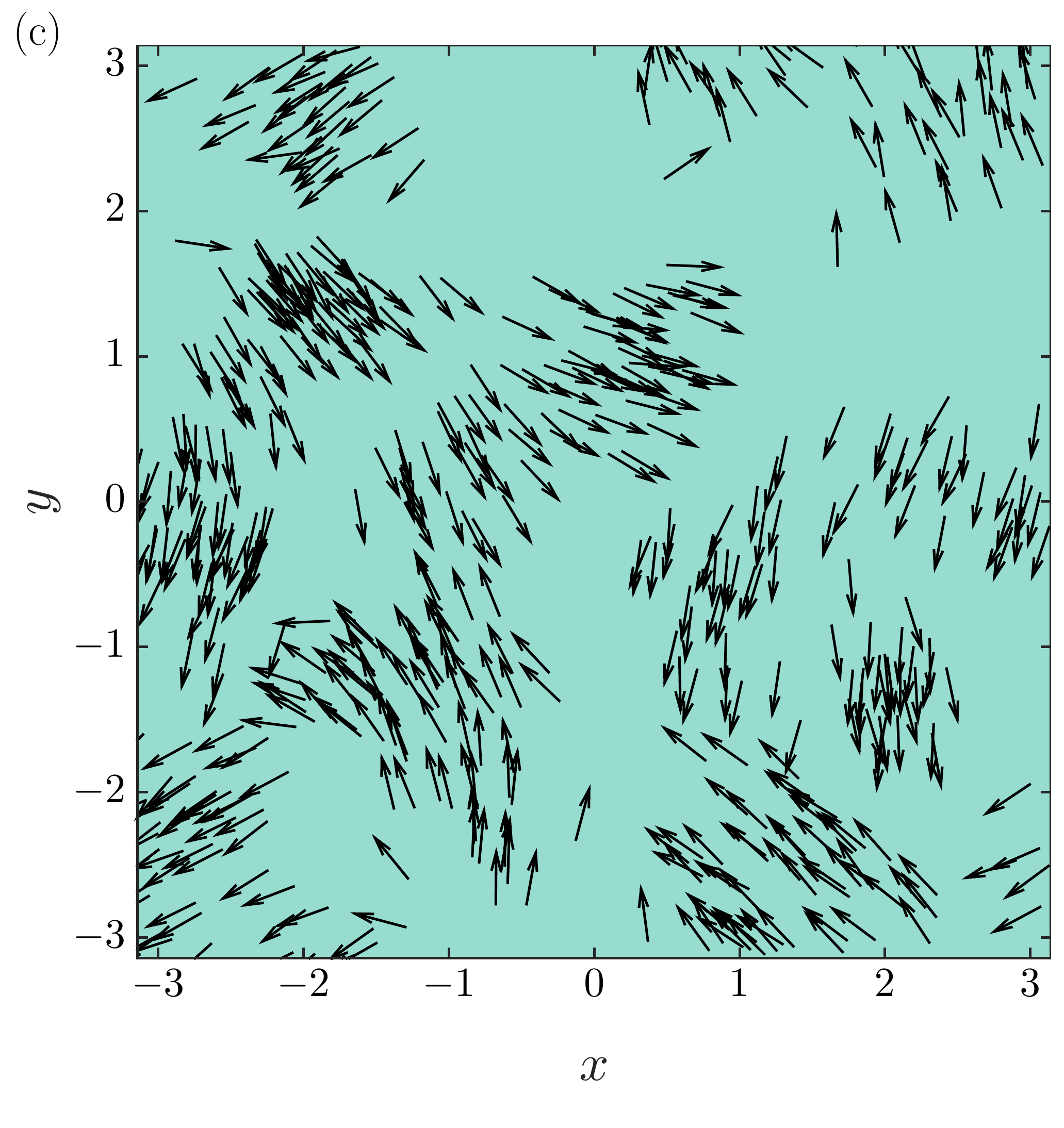}\hspace{5mm}
\includegraphics[width = 0.4\textwidth]{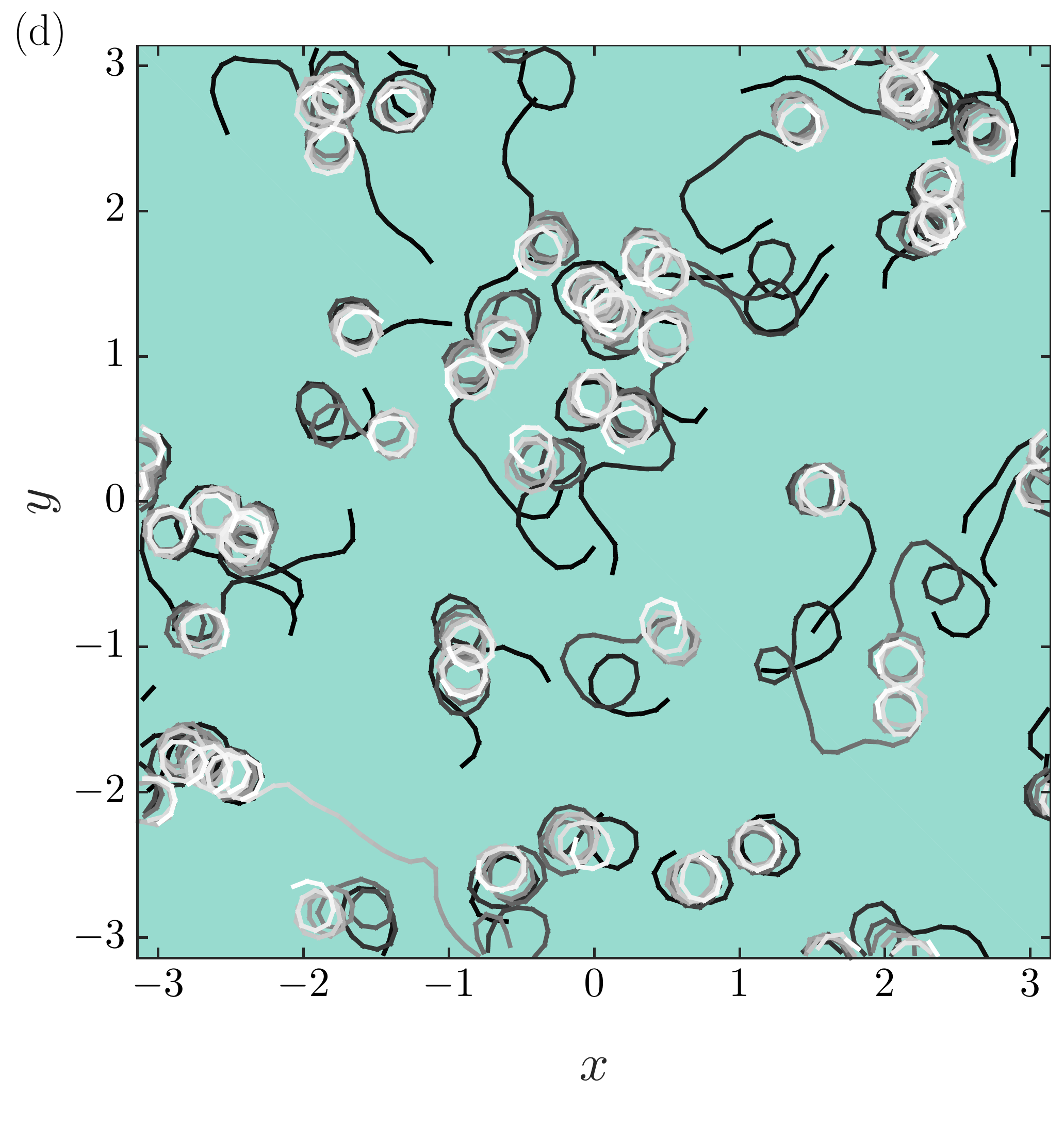}
\end{center}
\caption{{\bf Typical structure and trajectories of the flocks with no external flow.} (Left) Snapshots of the system in a statistically steady-state with radius of interaction $R=1.0$ ((a) $\alpha=0$, (c) $\alpha=\pi/2$); arrows indicate the particles direction of motion. (Right) corresponding particle trajectories(for 50 particles), dark to light indicates the direction of time.}
\label{fig:1}
\end{figure*}
%----------------------------------------------------------------------------------------------------------------------------------------
\section*{Modelling and Computational Methods}
We consider an extension to the self-propelled particle (SPP) model presented in \cite{Morin2015}, taking $N=500$ self-propelled particles in a two-dimensional square periodic domain with sides of size $L=2\pi$. Each particle has a position $\bx_i(t)$ and an intrinsic, self-driven, velocity $\bv_i(t)$. As is typical in SPP models, all particles are assumed to move with the same speed, $V=1$, and a particles intrinsic velocity is determined by
\begin{equation}\label{Eq:velocity}
\bv_i=(V\cos\theta_i,V\sin\theta_i),
\end{equation}
where $\theta_i$ determines the direction the particle moves in. In the Vicsek model \cite{Vicsek_1995} $\theta_i$ is periodically (at each time increment) determined from the average of the particle's own direction, plus the directions of its neighbours within a critical radius, $R$, such that
\begin{equation}\label{Eq:theta}
\theta_i=\big\langle \theta_j \big\rangle_{|\bx_i-\bx_j|< R} +\eta \xi_i,
\end{equation}
where angled brackets denote suitable averaging of the orientation of neighbours within the critical radius.
The final term in  Eq.~(\ref{Eq:theta}) is a noise term; specifically $\xi_i$ is a uniformly distributed random variable on the interval $[-1,1]$ and $\eta$ is the intensity of the noise. 

 Morin \etal \cite{Morin2015} proposed an extension to the model by including both alignment and anticipation such that the rate of change of orientation is given by
 \begin{align}
\dot \theta_i(t)=-\frac{1}{\tau}\big\langle\sin\left[{\theta_i-\left( \theta_j+\alpha \chi_j\right )}\right]\big\rangle_{|\bx_i-\bx_j|< R}+\eta \xi_i,
\label{eqtheta}
\end{align}
where $\alpha$ is a parameter we shall discuss shortly, $\chi_j\equiv\dot\theta_j/|\dot\theta_j|$ is the sign of the angular velocity (the particles spin) and $\tau$ is an orientation rate. This is more easily understood if we expand the sine function and recast Eq.~(\ref{eqtheta}) as:
 \begin{align}
\dot \theta_i(t)=&-\frac{1}{\tau}\cos\alpha\,\big\langle\sin\left({\theta_i-\theta_j}\right)\big\rangle_{|\bx_i-\bx_j|< R}
\label{vsalpha}
\\
&-\frac{1}{\tau}\sin\alpha\,\big\langle\sin\left[{\theta_i- (\theta_j+\chi_j\frac{\pi}{2})}\right]\big\rangle_{|\bx_i-\bx_j|< R}+\eta \xi_i.\nonumber
\end{align}
It is then clear that the first term is the standard Vicsek interaction which acts to promote alignment with orientations, whereas the second term promotes alignment with the acceleration of particles within the critical radius. The relative contribution of these two terms is determined by $\alpha$; in the limit $\tau \rightarrow 0$ with $\alpha=0$  we recover the Vicsek model. In contrast with $\alpha=\pi/2$ particles align purely with neighbouring particles' acceleration. It is worth noting that a model which included both alignment and anticipation of others motion was earlier considered in Szab\'o \etal \cite{Szabo2009} and we shall discuss their findings alongside our own later in the article. Morin \etal \cite{Morin2015}  showed that (contrary to what one might expect) including anticipation in the model does not enhance the stability of the flock. Indeed they found with increasing values of $\alpha$ there was a transition from a flocking state to a spinning state. This can be seen in Fig.~\ref{fig:1} where we plot snapshots of the system and particle trajectories.

\begin{figure*}[!htbp]
\begin{center}
\includegraphics[width = 0.375\textwidth]{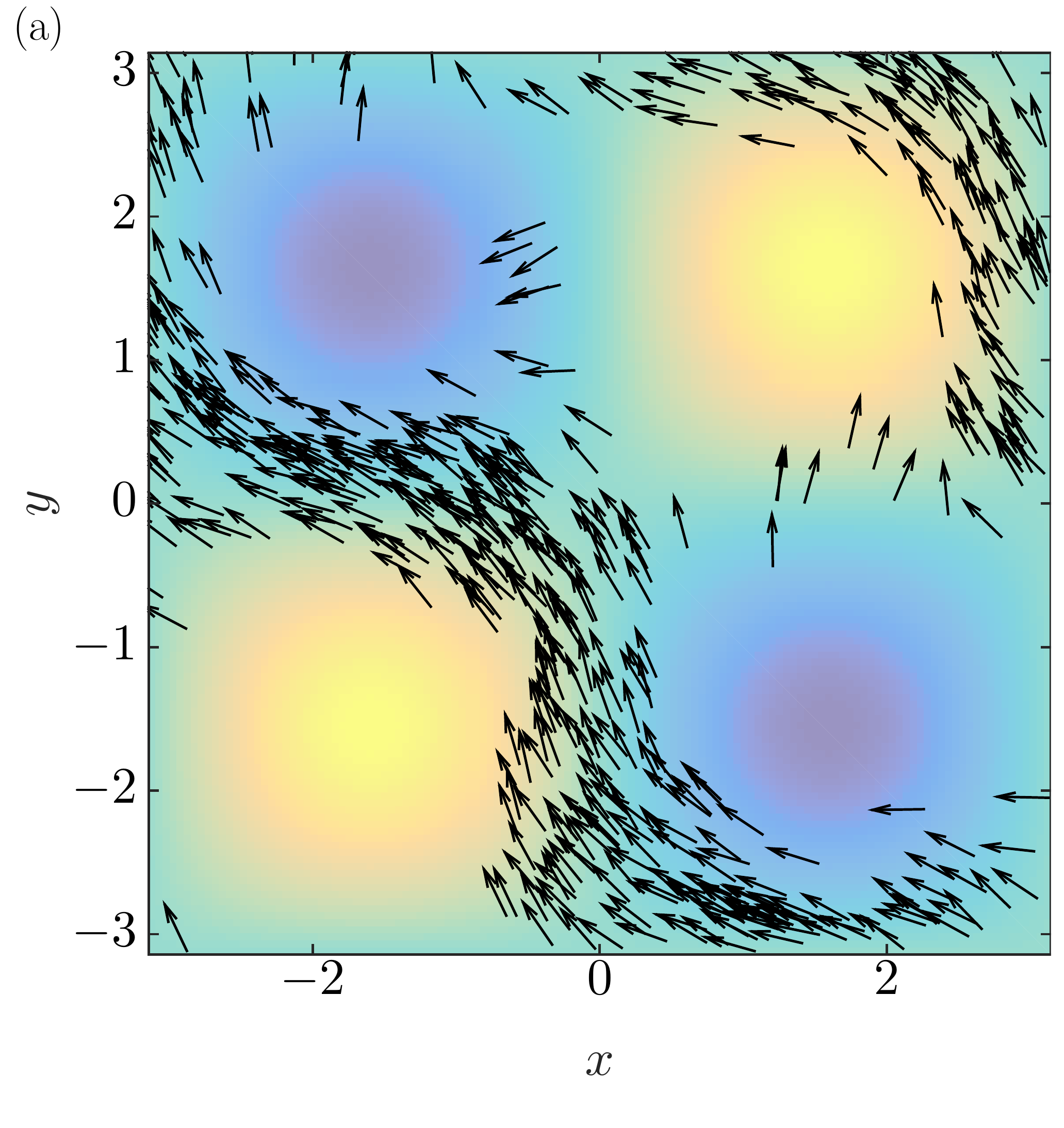}\hspace{5mm}
\includegraphics[width = 0.375\textwidth]{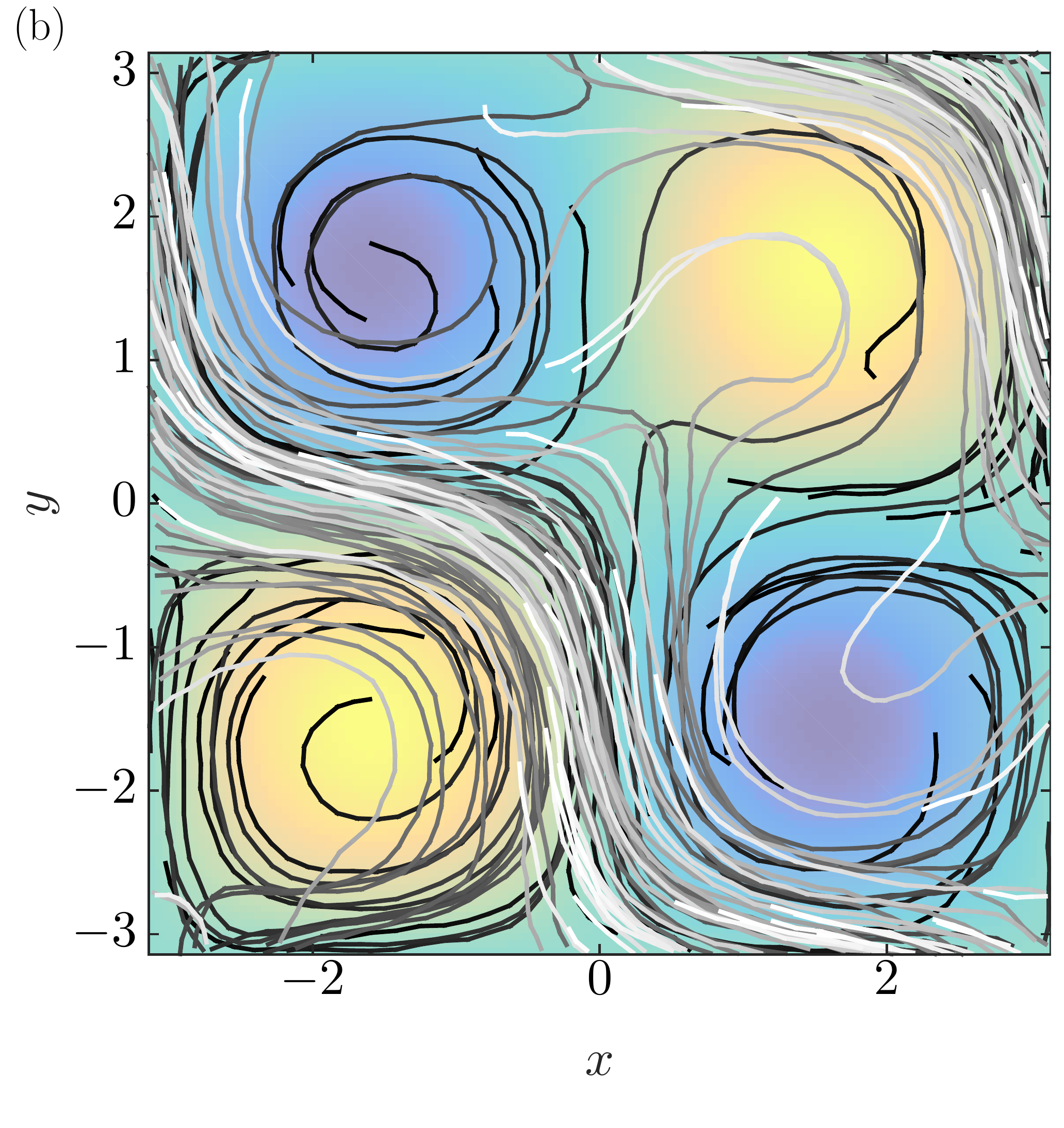}\\
\includegraphics[width = 0.375\textwidth]{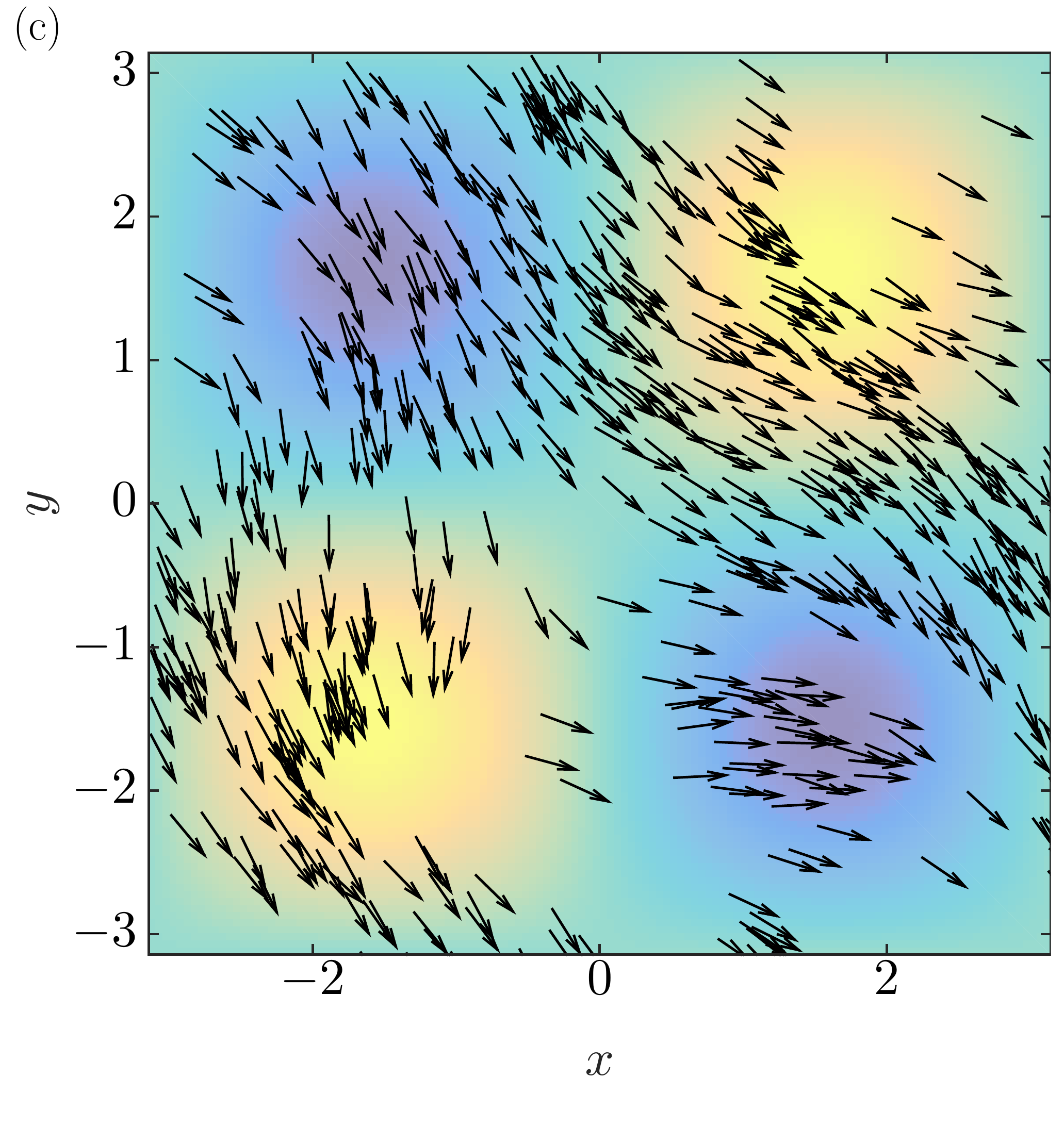}\hspace{5mm}
\includegraphics[width = 0.375\textwidth]{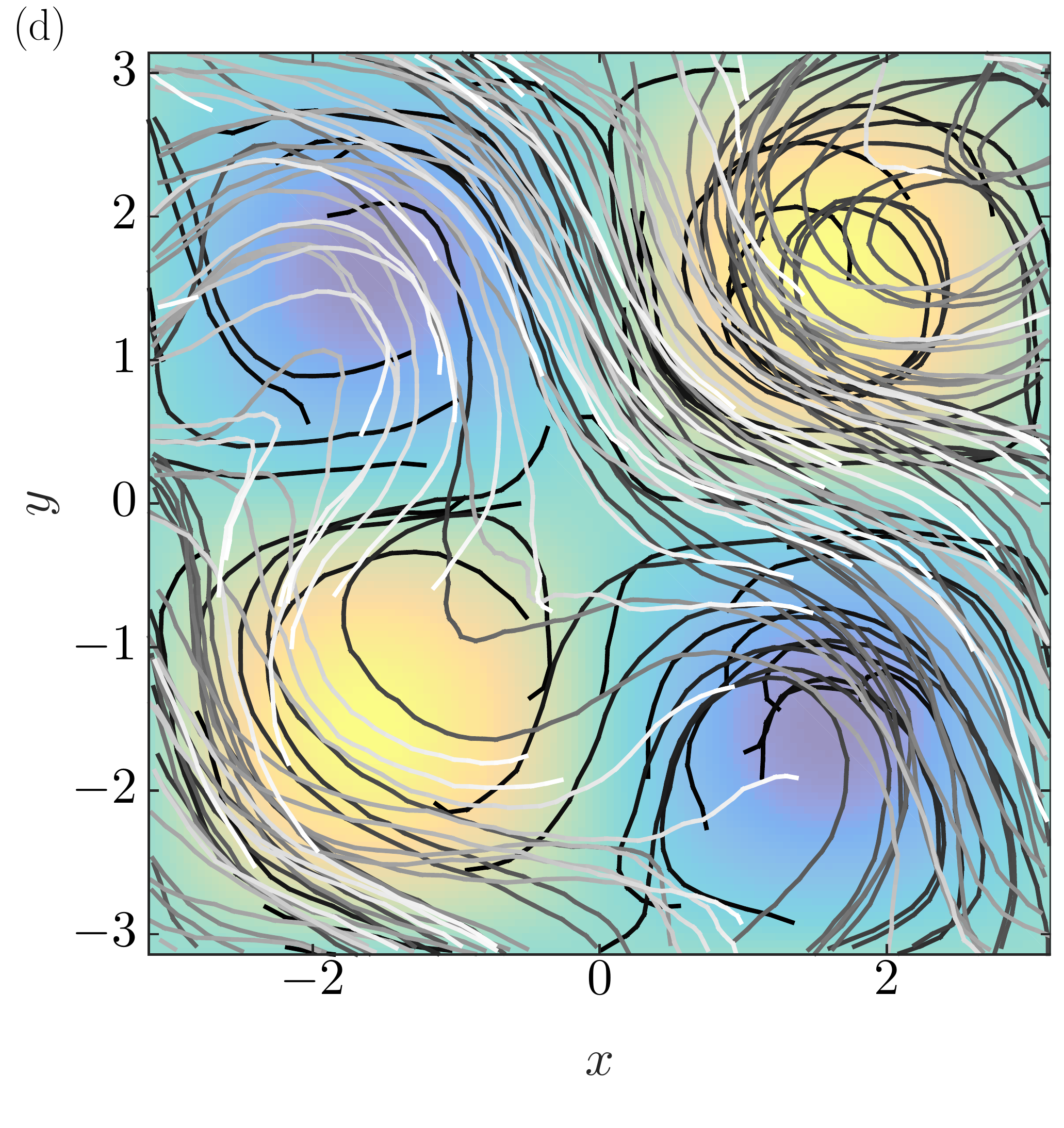}\\
\includegraphics[width = 0.375\textwidth]{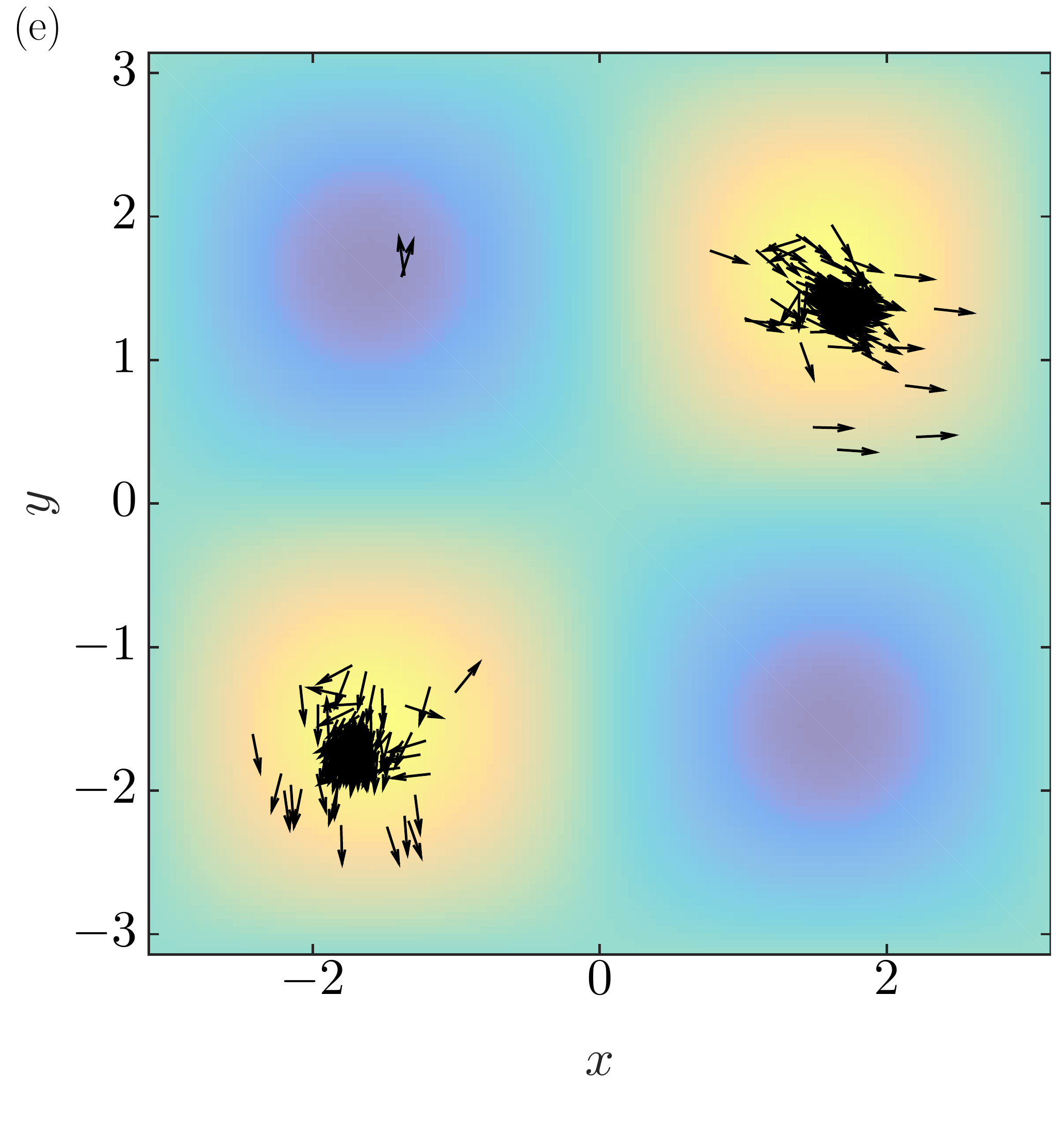}\hspace{5mm}
\includegraphics[width = 0.375\textwidth]{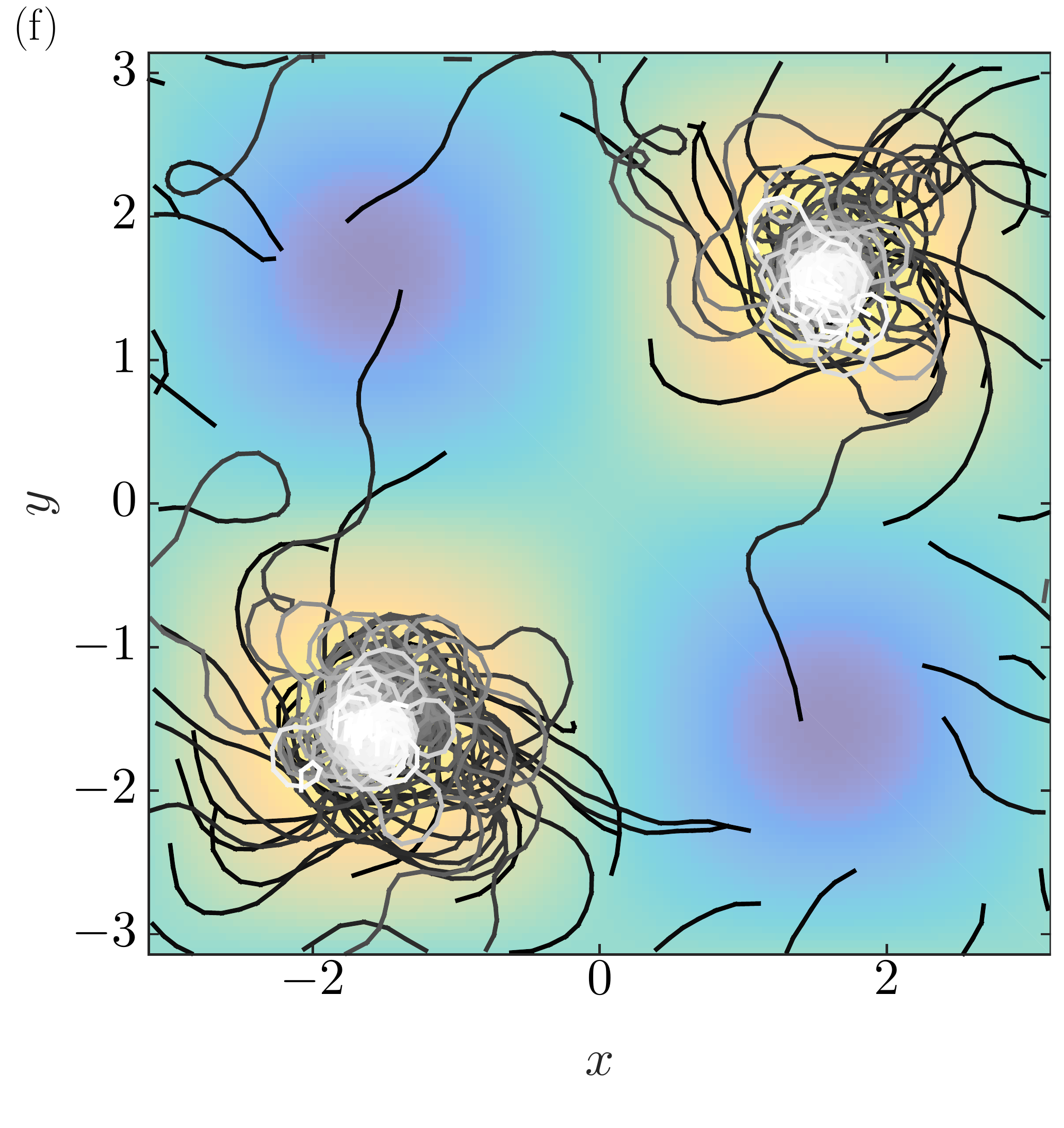}\\
\caption{{\bf Typical structure and trajectories of the flocks with in the Taylor-Green Vortex.} (Left) Snapshots of the system in a statistically steady-state with radius of interaction $R=1.0$ ((a) $\alpha=0$; (c)  $\alpha=0.4$;  (e) $\alpha=\pi/4$); $V_f=0.75$; arrows indicate the particles direction of motion. The magnitude of the flow vorticity is indicated by the pseudocolour plot, with light (yellow) corresponding to regions of large positive vorticity, and dark (blue) negative vorticity. (Right) corresponding particle trajectories (for 50 particles), dark to light indicates the direction of time.}
\label{fig:3}
\end{center}
\end{figure*}

In this paper we shall investigate if including anticipation can stabilise the flock in the presence of external noise which exhibits complex spatiotemporal correlations, such as one would expect flocks forming in a turbulent fluid environment would experience.
 Based on our previous arguments \cite{Baggaley2015} we forgo the computational expense and complexity of a direct numerical simulation (DNS) of the Navier-Stokes equations. Instead we turn to a well studied and widely used \cite{Durham2011,Maxey1986, Bergougnoux2014,Nore1997} `toy' flow, the Taylor Green (TG) vortex \cite{Taylor1923}, defined as 
\begin{equation}\label{eq:TGV}
\bvf(\bx)=(u_{f},v_{f})=V_f (\sin (x)\cos (y),-\cos (x)\sin (y)),
\end{equation}
where $\bx=(x,y)$. The vorticity field is given by 
\begin{equation}\label{eq:TGVw}
\omega=\nabla \times \bvf=2V_f\sin (x)\sin (y),
\end{equation}
the flow is incompressible ($\nabla \cdot \bvf=0$), and consists of cells of counter-rotating vortices as seen in Fig.~\ref{fig:3}.
$V_f$ is a scaling parameter which can be adjusted to modify the relative intrinsic particle speed to that of the background flow. The equation of motion for the SPPs is modified to
\begin{equation}\label{Eq:velocity2}
\dfrac{{\rm d}\bx_i}{{\rm d} t}=\mathbf{v}_i=(V\cos(\theta_i)+u_{f}(\bx_i),V\sin(\theta_i)+v_{f}(\bx_i)).
\end{equation}
We follow \cite{Khurana2013,Baggaley2015} and assume that particles orient themselves to the direction of motion of nearby particles, replacing $\theta_j$ in Eq.~(\ref{eqtheta}) with \cite{atan2}
\begin{equation}\label{eq:newtheta}
\breve{\theta}_j={\rm atan2}(V\sin(\theta_j)+v_{f}(\bx_j)),V\cos(\theta_j)+u_{f}(\bx_j)).
\end{equation} 
We retain the intrinsic noise ($\eta$) in Eq.~(\ref{eqtheta}), to model the fact that it is unlikely real animals will perfectly align themselves with neighbours within the critical radius and fix $\eta=0.2/\tau$. 

Particles are evolved according to an explicit Euler scheme such that
 \begin{align}
\bx_i(t+\Delta t)&=\bx_i(t)+\Delta t \bv_i(t)\\
\theta_i(t+\Delta t)&=\theta_i(t)+\Delta t \dot \theta_i(t) \nonumber
 \end{align}
where at each timestep $\bv_i(t)$ is updated according to Eqns.~(\ref{Eq:velocity2}), (\ref{eq:newtheta}) \& (\ref{eqtheta}). Note $\xi_i$ is drawn randomly at each timestep, we take $\Delta t=0.05$ and in each simulation evolve the system for  $2\times 10^3$ timesteps. We fix $\tau=\Delta t$, such that if $\alpha=0$ we recover the standard Vicsek model. The global order of the system can be characterised by computing,
\begin{equation}
\psi(t)=\dfrac{1}{NV}\left | \sum_{i=1}^N \bv_i \right |.
\end{equation}
In a typical simulation $\psi$ grows from $0$ until it saturates and fluctuates around some mean value, which depends on $R$, $\alpha$ and $V_f$, hence it is a convenient measure to establish if the system has reached a statistically steady state. As we are interested in systems which exhibit flocking behaviour in the limit of small intrinsic noise and no extrinsic noise, we set $R=1.0$ for all simulations reported here.

We then perform a suite of numerical simulations to thoroughly investigate a two dimensional $(\alpha,V_f)$-parameter space, with $\alpha \in [0,\pi/4]$ and $V_f\in[0.1,1.25]$. For each point in $(\alpha,V_f)$-space we perform $i=1,\ldots,10$ simulations, computing the mean value of $\psi_{\alpha,V_f,i}$ and its variance $\sigma^2_{\alpha,V_f,i}$ in each simulation (once it has reached a statistically steady value).  We report the ensemble averaged mean (weighted by the inverse of the variance) value over the 10 simulations. We denote this value $\langle \psi \rangle$, where the angled brackets indicate the use of temporal and ensemble averaging, by taking a weighted mean the standard deviation of $\langle \psi \rangle$ (for a given $\alpha$ and $V_f$) is \cite{eadie1971statistical}
\begin{equation}\label{eq:sigma}
\sigma_{\langle \psi \rangle}=\sqrt{\left ( \sum_{i=1}^n \sigma_i^{-2} \right )^{-1}}.
\end{equation}

 %----------------------------------------------------------------------------------------------------------------------------------------
\begin{figure}[h]
\begin{center}
\includegraphics[width = 0.45\textwidth]{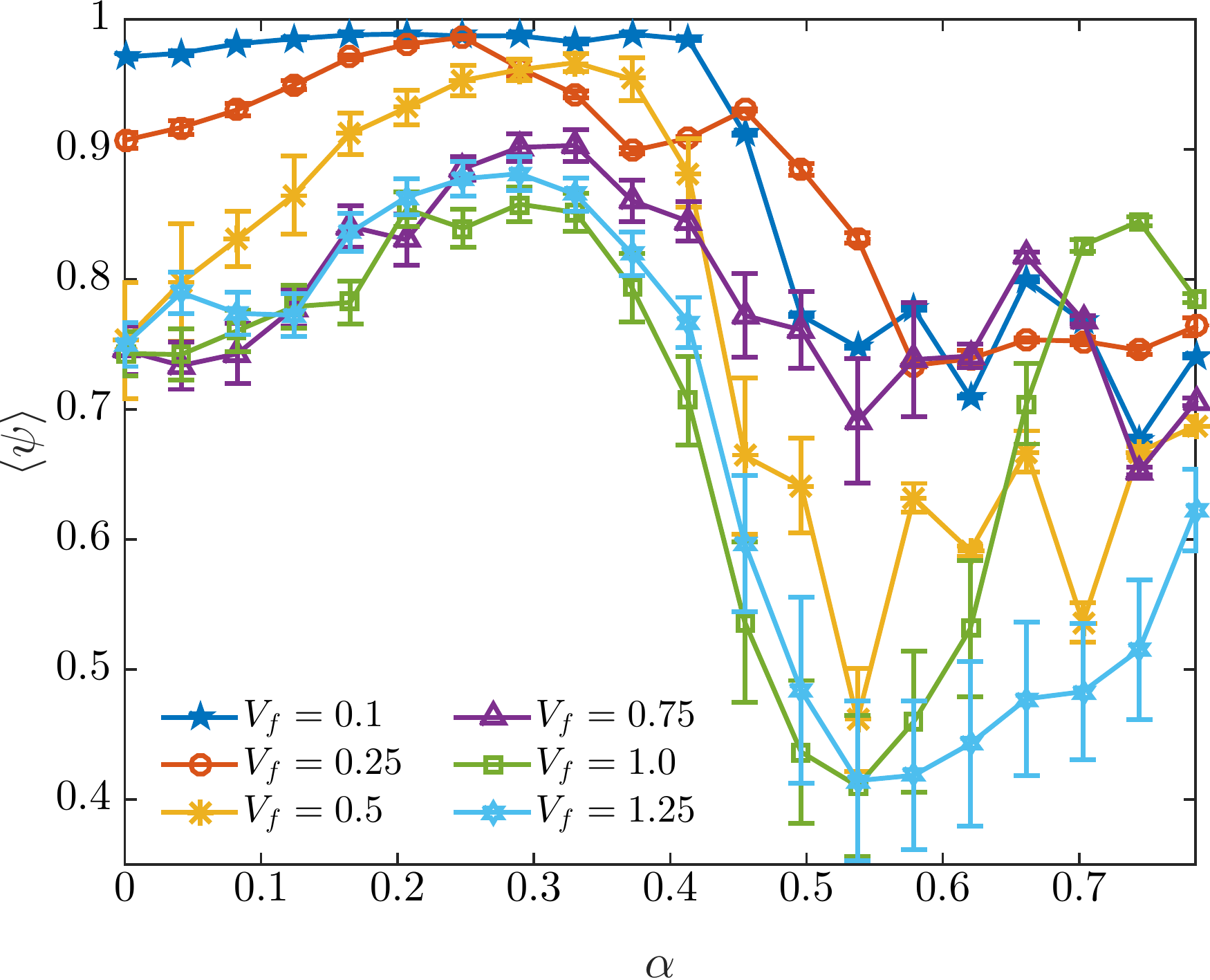}
\caption{{\bf Flock stability.} Temporally and ensemble averaged (denoted by angled brackets) values of the order parameter $\psi$, plotted as a function $\alpha$ (see Eq.~(\ref{eqtheta}) \& (\ref{vsalpha})), for varying flow speed $V_f$. Errorbars are given by $\pm 2\sigma_{\langle \psi \rangle}$, see Eq.~(\ref{eq:sigma}).}
\label{fig:2}
\end{center}
\end{figure}
%----------------------------------------------------------------------------------------------------------------------------------------
%----------------------------------------------------------------------------------------------------------------------------------------

 %----------------------------------------------------------------------------------------------------------------------------------------
\begin{figure}[h]
\begin{center}
\includegraphics[width = 0.45\textwidth]{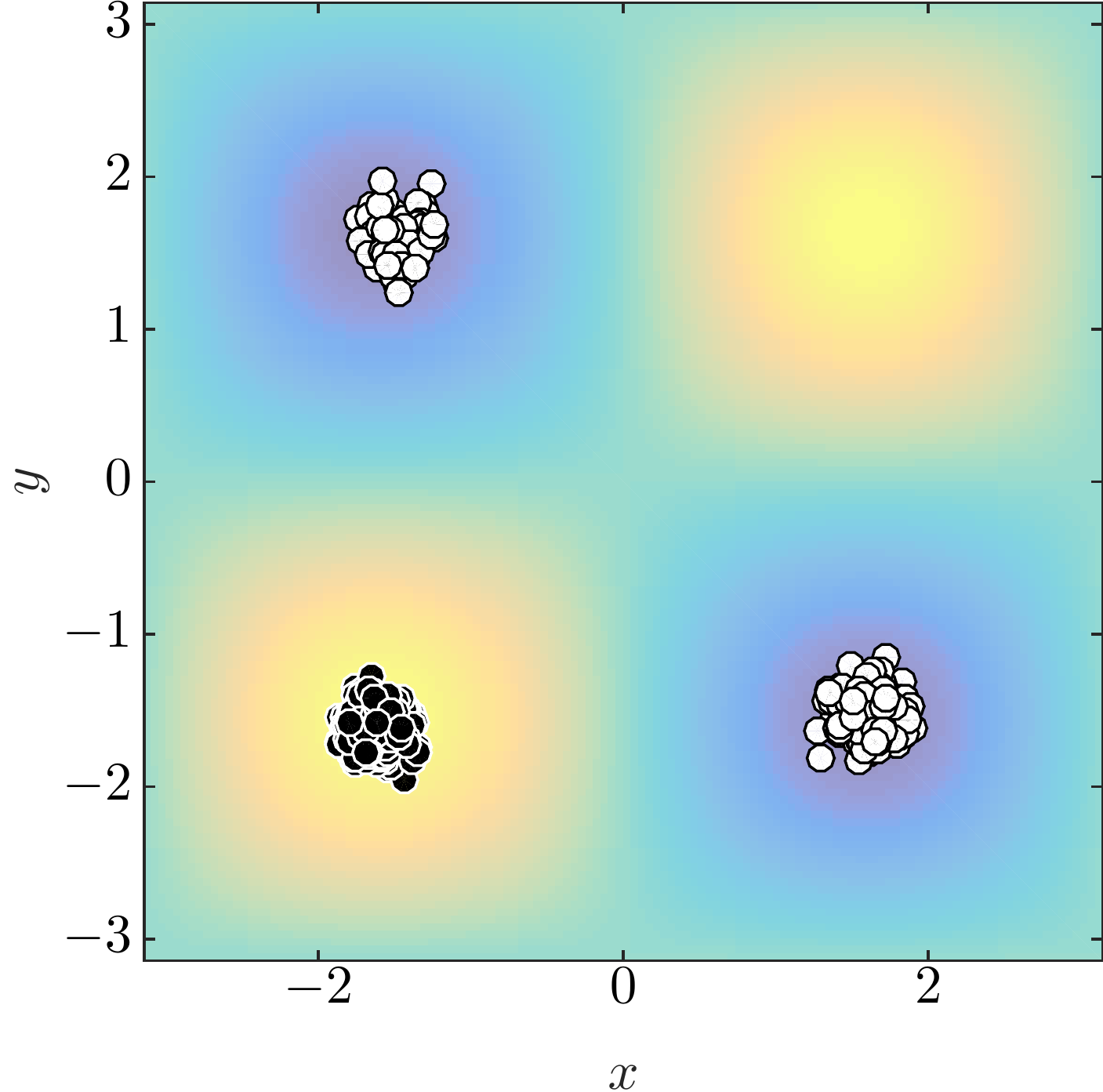}
\caption{{\bf Spin-vorticity matching.} Snapshots of the system in a statistically steady-state with $R=1.0$, $\alpha=\pi/4$ and $V_f=0.75$. Particles with positive spin ( $\chi_i\equiv\dot\theta_i/|\dot\theta_i|=1$ ) are plotted as black circles, those with negative spin ($\chi_i=-1$) as white circles.  The magnitude of the flow vorticity is indicated by the pseudocolour plot, with light (yellow) corresponding to regions of large positive vorticity, and dark (blue) negative vorticity. }
\label{fig:4}
\end{center}
\end{figure}
%----------------------------------------------------------------------------------------------------------------------------------------

%----------------------------------------------------------------------------------------------------------------------------------------

\section*{Results}
Our main results are presented in Fig.~\ref{fig:2} where we plot $\langle \psi \rangle$ vs. $\alpha$ for varying $V_f$. For all values of $V_f$ a moderate value of $\alpha$ is seen to enhance the global alignment of the the flock, at larger values of $\alpha$ the stability breaks down, as particles form smaller clusters which follow tight spiral trajectories. However what is striking is that as the flow speed increases anticipation is seen to have a profound stabilising effect. Note also that there is a reduction in the value of $\sigma_{\langle \psi \rangle}$ for moderate values of $\alpha$, at least for $V_f<1.0$, which indicates a reduction in the magnitude of the fluctuations of $\psi$. One would imagine that this is also advantageous allowing information (e.g. changes in direction, arrival of a predator) to propagate more efficiently through the flock.

%----------------------------------------------------------------------------------------------------------------------------------------
\begin{figure*}[!t]
\begin{center}
\includegraphics[width = 0.45\textwidth]{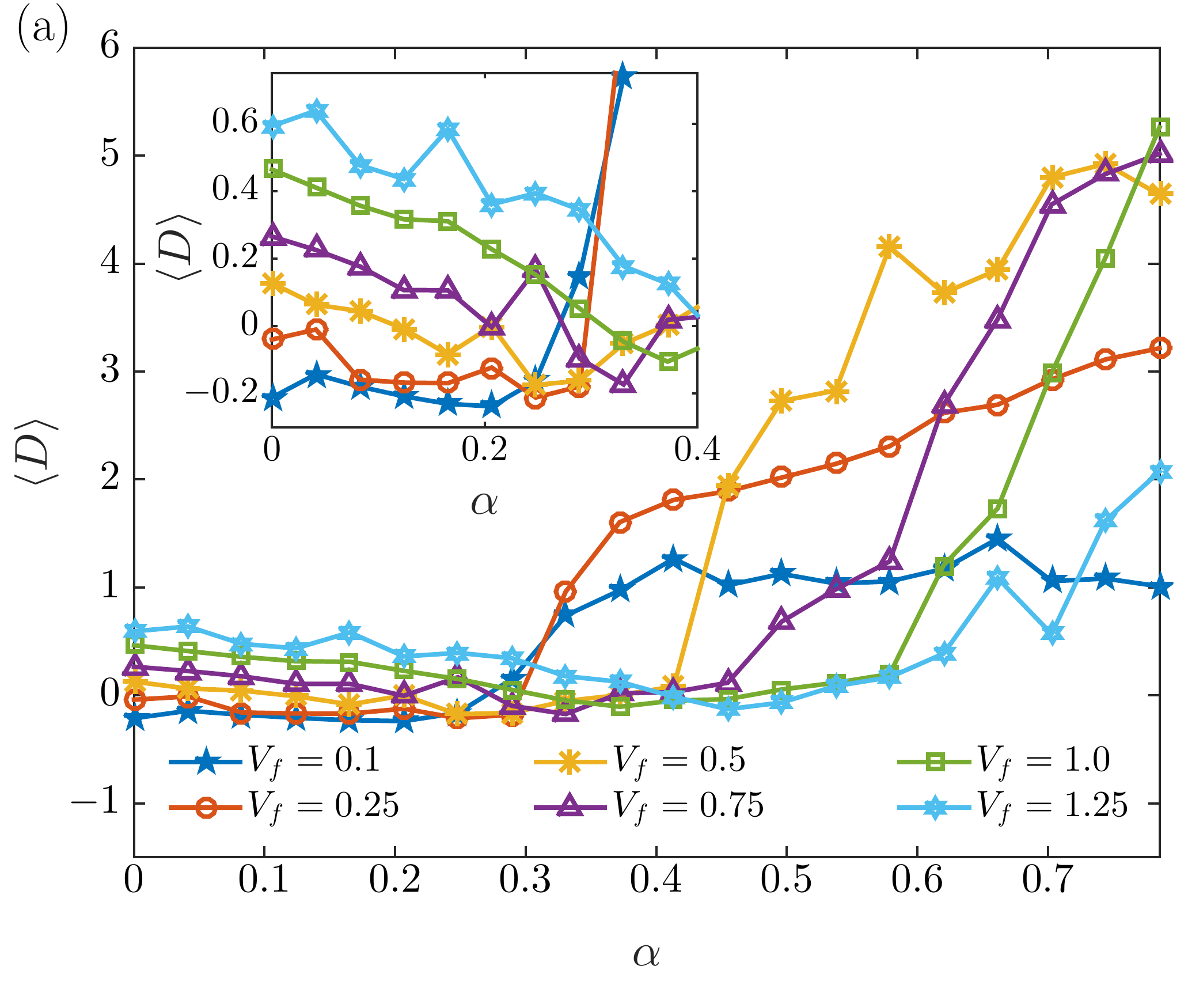}\hspace{7mm}
\includegraphics[width = 0.45\textwidth]{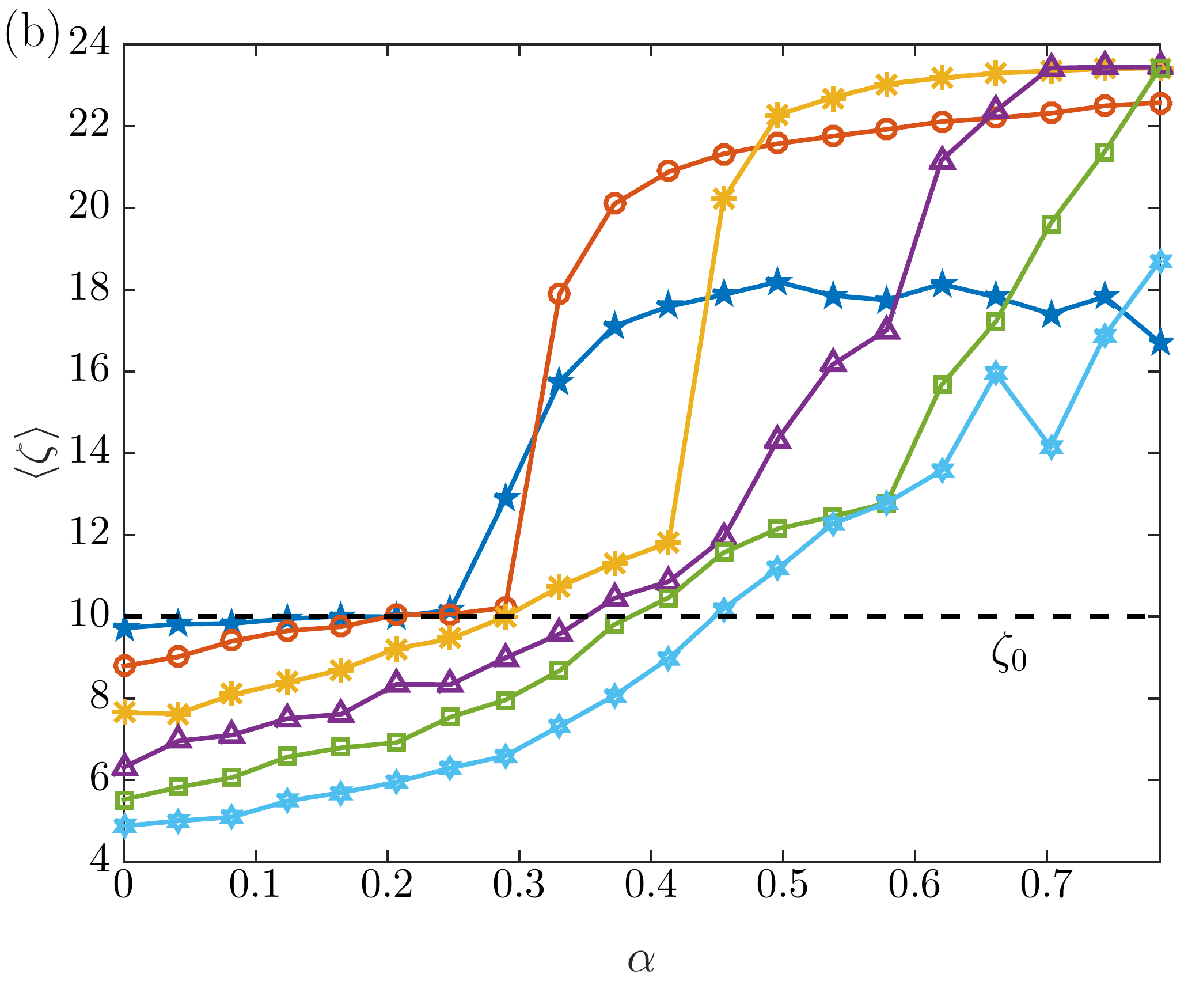}
\caption{{\bf Spatial distribution of the particles:} (a) The `patchiness' of the spatial distribution of points. Large values of $D$ indicate the particles are confined to a small region of the domain, $D=0$ indicates a random distribution of particles, and $D<0$ indicates segregation of particles. The inset shows a zoom in of the region where $\alpha<0.4$. (b) $\zeta$ (E.q.~\ref{Eq:zeta}) plotted as a function of $\alpha$, the dashed line indicates a random distribution of points; $\zeta<0$ particles confined to regions of low vorticity; $\zeta>0$ particles confined to regions of high vorticity.}
\label{fig:5}
\end{center}
\end{figure*}
%----------------------------------------------------------------------------------------------------------------------------------------
In order to understand this phenomena in Fig. ~\ref{fig:3} we plot particle trajectories and snapshots of the system for $V_f=0.75$ with varying $\alpha$. For $\alpha=0$, i.e. the Vicsek model, we see the particles are expelled from regions of high vorticity, and form filamentary structures as reported in our earlier work \cite{Baggaley2015}. In contrast for large values of $\alpha$, where anticipation becomes dominant particles move into the areas of high vorticity and form small clusters where each particle follows a tight spiral trajectory. However at the interface of these two regimes we find that the `correct' amount of anticipation can counteract the destabilising effect of the imposed flow field. 
We note that in the large $\alpha$ limit, where particles move into areas of high vorticity there is a `matching' between the particles spin, i.e. the sign of its angular velocity, and the sign of the vorticity, as is clear in Fig.\ref{fig:4}.

Interestingly in Szab\'o \etal \cite{Szabo2009} they found that the information exchange between particles was maximised at a critical balance between alignment and anticipation. They conjectured (due to the importance of information exchange in animal societies) that such a critical balance may provide an optimal behavioural strategy. Here we show that it also provides a method to overcome the destabilising effects of spatially correlated noise.

In order to quantify the results presented in Fig.~\ref{fig:3} we define two relevant statistics. Firstly we quantify 
the `patchiness' of the spatial distribution of particles in the domain. Following \cite{Fessler1994, Durham2011} we course-graining the particles onto a 16 by 16 regular array of boxes. Within each box we compute the particle density (based on the number of particles lying within the box) and denote this quantity $n(\bx,t)$. As the particle density in each box is Poisson distributed this has a mean value $E[n]=\lambda=N/4\pi^2\simeq12.6$. If particles preferentially accumulate in certain regions of the domain the standard deviation of $n$, $\sigma_n$, increases relative to its initial value, $\sigma_P=\lambda^{1/2}$. Hence $\sigma_n$ can be appropriately normalised to give the accumulation index \cite{Fessler1994} $D=(\sigma_n-\sigma_P)/\lambda$, which is a measure of the spatial distribution of the points in the domain. Large values of $D$ indicate patchiness, i.e. the particles are concentrated in smaller subdomain(s), $D=0$ indicates a random distribution of particles, and $D<0$ indicates segregation of particles, relative to a random distribution.

To extract the regions the particles are located, we use define $\zeta$ to be 
\begin{equation}\label{Eq:zeta}
\zeta=\int_{\cal A} n |\boldsymbol{\omega}| dA,
\end{equation}
the integral of the product of the particle density field and the modulus of the flow's vorticity field. For a random distribution of particles we would expect $\zeta=\zeta_0=\bar{\boldsymbol{|\omega|}}(N/4\pi^2) \simeq 10$, where the overbar denotes the spatial mean of the modulus of the vorticity field. If the particles are concentrated in regions of vanishing vorticity then we would expect $\zeta \simeq 0$. Conversely if $\zeta>\zeta_0$ then particles are concentrated in regions of high vorticity.

Figure \ref{fig:5} shows the temporally and ensemble averaged (as described above) values of $D$ (left panel) and  $\zeta$ (right). We see without any anticipation, as the flow speed increases particles are confined into the regions of low vorticity, hence increasing values of $\langle D \rangle$, and decreasing values of $\langle \zeta \rangle$. However some anticipation (the optimal amount depending on the flow speed as one may expect) is seen to lead to values of $\zeta \sim \zeta_0$. Finally we see for large values of $\alpha$ the particles tend to collect in regions of high vorticity consistent with our earlier discussion of Fig.~\ref{fig:3}.

This also ties into our earlier discussion about the motivation for collective motion, in terms of safety in numbers to minimise the `domain of danger'. Clearly with too little anticipation (where the particles in our model are forced into thin filamentary structures) or too much (where particles concentrate in dense patches) the morphology of the flock is not optimal for providing increased safety in numbers. However a balance between these two competing affects does appear at least one viable strategy for the flock's morphology to not be strongly influenced by the underlying structure of the external fluid forcing.

\section*{Summary}
To summarise we have investigated an extension to the widely used Vicsek model in which collective motion emerges due to alignment with neighbouring particles and anticipation of their motion. With the addition of extrinsic noise in the form of a steady vortical flow we find the global order of the flock is significantly reduced and particles are confined to regions of low vorticity. In contrast in a model based purely on anticipation we find particles concentrate in regions of high vorticity. Most strikingly we find particles with a critical balance of alignment and anticipation are no longer slave to the flow, and global coherence emerges. At this critical balance (for $V_f<1.0$) we also see a reduction in the magnitude of the fluctuations of $\psi$, which surely would also be advantageous to members of the flock.

Hence one strategy for animals flocking in a complex (i.e. turbulent) flow could be not only align with neighbours but also to anticipate their motion, which seems entirely plausible. In addition our findings could have implications for flocking autonomous drones (unmanned ariel vehicles) and artificial microswimmers \cite{Dreyfus:2005aa}. By varying the amount of alignment and anticipation different regions of a fluid could be probed, or by tuning their relative contributions the separation between devices could be maximised, i.e. to prevent collisions. Whilst this clearly does not mark the end of the story, particularly in biological systems, we strongly believe that by studying how flocks react to external perturbations (fluid motion, predatory threats etc.) and comparing to the dynamics of models will enhance our understanding of collective motion in biological systems.


\begin{thebibliography}{10}
\bibitem{atan2}
Note the use of the atan2 function, the four-quadrant inverse tangent
  $\tan^{-1}(y,x)$. Two arguments are required, so that the signs of the inputs
  is not lost, essential to return the correct quadrant of the computed angle.
\end{thebibliography}
\end{document}